\documentclass[twocolumn, preprintnumbers,prb,aps,amssymb,showpacs,superscriptaddress]{revtex4}
\usepackage{graphicx}
\usepackage{dcolumn}
\usepackage{wasysym}
\usepackage{bm} \usepackage{color}
\begin{document}

\newcommand{\KBa}{Ba$_{1-x}$K$_{x}$Fe$_2$As$_2$}
\newcommand{\TN}{$T_{\rm N}$}
\newcommand{\Tc}{$T_{\rm c}$}
\newcommand{\To}{$T_0$}
\newcommand{\ie}{{\it i.e.}}
\newcommand{\eg}{{\it e.g.}}
\newcommand{\etal}{{\it et al.}}
\newcommand{\K}{Ba$_{1-x}$K$_x$Fe$_2$As$_2$}
\newcommand{\KFeAs}{KFe$_2$As$_2$}
\newcommand{\Co}{Ba(Fe$_{1-x}$Co$_x$)$_2$As$_2$}
\newcommand{\Kzero}{$\kappa_0/T$}
\newcommand{\units}{$\mu \text{W}/\text{K}^2\text{cm}$}
\newcommand{\p}[1]{\left( #1 \right)}
\newcommand{\Dd}[2]{\frac{\text{d} #1}{\text{d}#2}}


\title{
Expansion of the tetragonal magnetic phase with pressure in the iron-arsenide
superconductor \KBa
}


\author{E.~Hassinger}
\email{elena.hassinger@usherbrooke.ca}
\affiliation{D\'epartement de physique \& RQMP, Universit\'e de Sherbrooke, Sherbrooke, Qu\'ebec J1K 2R1, Canada}

\author{G.~Gredat}
\affiliation{D\'epartement de physique \& RQMP, Universit\'e de Sherbrooke, Sherbrooke, Qu\'ebec J1K 2R1, Canada}

\author{F.~Valade}
\affiliation{D\'epartement de physique \& RQMP, Universit\'e de Sherbrooke, Sherbrooke, Qu\'ebec J1K 2R1, Canada}

\author{S.~Ren\'e~de~Cotret} 
\affiliation{D\'epartement de physique \& RQMP, Universit\'e de Sherbrooke, Sherbrooke, Qu\'ebec J1K 2R1, Canada}

\author{O.~Cyr-Choini\`ere} 
\affiliation{D\'epartement de physique \& RQMP, Universit\'e de Sherbrooke, Sherbrooke, Qu\'ebec J1K 2R1, Canada}

\author{A.~Juneau-Fecteau} 
\affiliation{D\'epartement de physique \& RQMP, Universit\'e de Sherbrooke, Sherbrooke, Qu\'ebec J1K 2R1, Canada}

\author{J.-Ph.~Reid}
\affiliation{D\'epartement de physique \& RQMP, Universit\'e de Sherbrooke, Sherbrooke, Qu\'ebec J1K 2R1, Canada}

\author{H.~Kim}
\affiliation{Ames Laboratory, Ames, Iowa 50011, USA}

\author{M.~A.~Tanatar}
\affiliation{Ames Laboratory, Ames, Iowa 50011, USA}

\author{R.~Prozorov}
\affiliation{Ames Laboratory, Ames, Iowa 50011, USA} 
\affiliation{Department of Physics and Astronomy, Iowa State University, Ames, Iowa 50011, USA }

\author{B.~Shen}
\affiliation{Center for Superconducting Physics and Materials, National Laboratory of Solid State Microstructures
and Department of Physics, Nanjing University, Nanjing 210093, China}

\author{H.-H.~Wen}
\affiliation{Center for Superconducting Physics and Materials, National Laboratory of Solid State Microstructures
and Department of Physics, Nanjing University, Nanjing 210093, China}
\affiliation{Canadian Institute for Advanced Research, Toronto, Ontario M5G 1Z8, Canada}

\author{N.~Doiron-Leyraud} 
\affiliation{D\'epartement de physique \& RQMP, Universit\'e de Sherbrooke, Sherbrooke, Qu\'ebec J1K 2R1, Canada}

\author{Louis Taillefer}
\email{louis.taillefer@usherbrooke.ca}
\affiliation{D\'epartement de physique \& RQMP, Universit\'e de Sherbrooke, Sherbrooke, Qu\'ebec J1K 2R1, Canada}
\affiliation{Canadian Institute for Advanced Research, Toronto, Ontario M5G 1Z8, Canada}

\date{\today}


\begin{abstract}
In the 
temperature-concentration phase diagram of 
most iron-based superconductors, antiferromagnetic order is gradually suppressed to zero at
a critical point, and a dome of superconductivity forms around that point.
The nature of the magnetic phase and its fluctuations is of fundamental importance for elucidating the pairing mechanism.
In \KBa~and Ba$_{1-x}$Na$_x$Fe$_2$As$_2$, it has recently become clear that the 
usual stripe-like magnetic phase, of orthorhombic symmetry,
gives way to a second magnetic phase, of tetragonal symmetry, near the critical point, between $x = 0.24$ and $x = 0.28$.
Here we report measurements of the electrical resistivity of \KBa~under applied hydrostatic pressures up to 2.75~GPa,
for $x = 0.22$, 0.24 and 0.28.
We track the onset of the tetragonal magnetic phase using the sharp anomaly it produces in the resistivity.
In the temperature-concentration phase diagram of \KBa,
we find that pressure greatly expands
the tetragonal magnetic phase,
while the stripe-like phase shrinks.
This raises the interesting possibility that the fluctuations of the former phase might be involved
in the pairing mechanism 
responsible for the superconductivity.


\end{abstract}


\maketitle


The phase diagram of iron-based superconductors of the BaFe$_2$As$_2$ family is characterized by competing antiferromagnetic (AF) order and superconductivity. 
Usually, the AF order decreases with 
concentration (doping)
and a dome of superconductivity surrounds the critical point.\citep{Canfield_ARCMP_2010} 
The AF order is a stripe-like spin-density wave, with a wavevector ${\bf Q} = (\pi, 0)$
and the magnetic moments lie in the plane. 
At the magnetic transition temperature, or slightly above it, the lattice changes from tetragonal at high temperature to orthorhombic at low temperature.\citep{Pratt_PRL_2009,Avci2012}

In Ba$_{1-x}$X$_x$Fe$_2$As$_2$, where X = K or Na, the phase diagram was recently found to be richer than this simple picture. 
Resistivity measurements under pressure revealed the existence of an internal transition inside the AF phase of \KBa.\citep{Hassinger_2012_PRB}
As the onset temperature \TN~of the orthorhombic AF phase (o-AF) 
is suppressed with hydrostatic pressure $P$, an additional phase transition to a ``new phase'' appears below a transition temperature \To~$<$~\TN, for $0.16 <�x < 0.21$, when $P > 0.9$~GPa.\citep{Hassinger_2012_PRB}
A tetragonal magnetic phase (t-AF) was then discovered 
in the closely related compound Ba$_{1-x}$Na$_x$Fe$_2$As$_2$, 
for $0.24 < x < 0.28$, 
by neutron and x-ray diffraction on powder samples.\citep{Avci_2014_NatCom}
Subsequent neutron scattering on single crystals showed that 
in this t-AF phase 
the spins are aligned parallel to the $c$ axis.\citep{Wasser_2015_PRB}
A similar phase of tetragonal symmetry was then found in \KBa~at ambient pressure, 
for $0.24 < x < 0.28$.\citep{Boehmer_2015_NatCom}
The magnetic moments in the t-AF phase of \KBa~are also oriented along the $c$~axis.\citep{Allred_2015_PRB,Mallet_2015_PRL}
Infrared spectroscopy showed that the t-AF phase has a double-$Q$ magnetic structure,\citep{Mallet_2015_PRL}
as opposed to the single-$Q$ structure of the o-AF phase.
Several theoretical studies have investigated the properties of the tetragonal magnetic phase in iron-based superconductors.
\citep{Lorenzana_2008_PRL,Eremin_PRB_2010,Berg_PRB_2010,Giovanetti_NatComm_2011,Brydon_PRB_2011,Avci_2014_NatCom,Wang_PRB_2015,Kang_PRB_2015,Fernandes_arXiv_2015,Gastiasoro_arXiv_2015,Osborn_arXiv_2015}

In this Article, we extend our prior study of \KBa~under pressure, performed up to $x~=~0.21$,\citep{Hassinger_2012_PRB}
by studying three further samples, with $x = 0.22$, 0.24 and 0.28. 
We are able to connect the additional phase induced by pressure with the tetragonal phase seen at ambient pressure. 
Pressure is seen to cause a dramatic expansion of the tetragonal magnetic phase, on the backdrop of a shrinking orthorhombic phase.
%

\begin{figure*}[t]
\centering
\includegraphics[width=18.0cm]{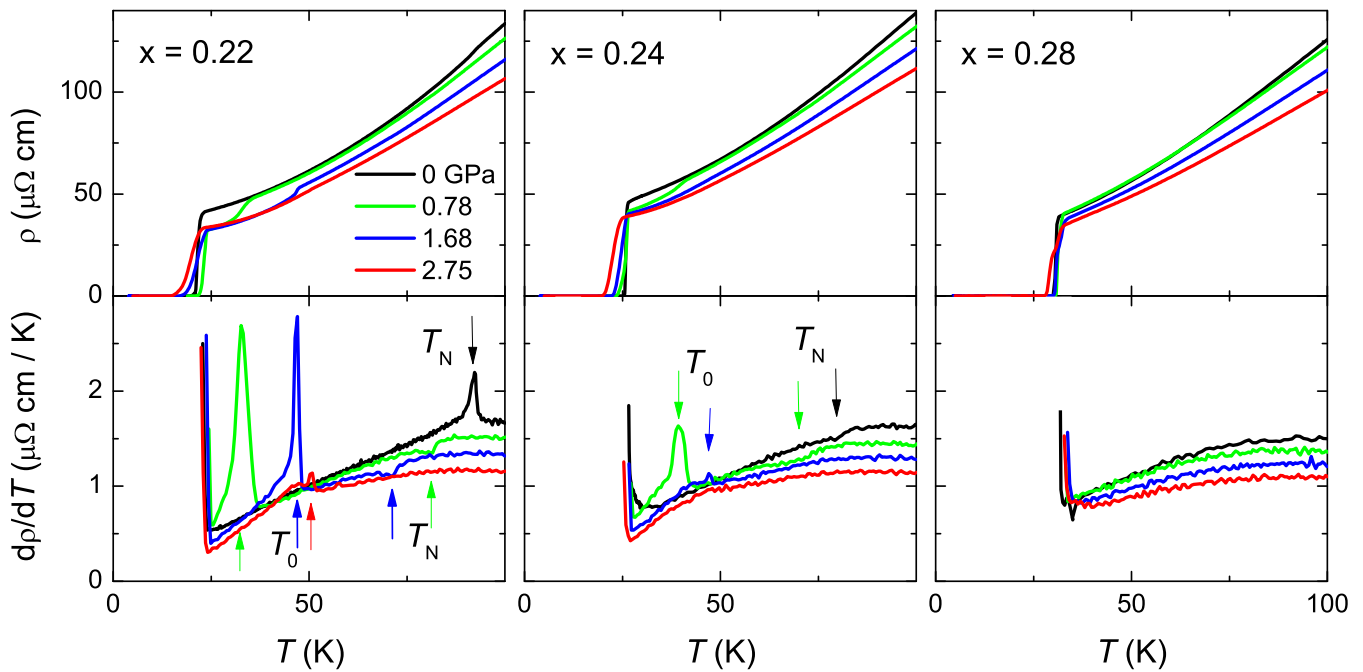}
\caption{
{\it Top}:
In-plane electrical resistivity of \KBa~for $x = 0.22$, $x = 0.24$ and $x = 0.28$ (different columns) for four different pressures, as indicated.
{\it Bottom}: 
Temperature derivative of the data in the top panels. 
The peak (dip) between 60~K and 100~K signals the onset of stripe-like antiferromagnetic order at \TN~(arrows).
The peak at lower temperature signals the onset of the tetragonal magnetic phase at \To~(arrows).
}
\label{300K}
\end{figure*}


{\it Methods.--}
Single crystals of \KBa~were grown from self flux.\citep{Growth}
Three underdoped samples were measured, with a superconducting transition temperature 
\Tc~$= 20.8 \pm 0.5$~K, $25.4 \pm 0.5$~K, and $30.1 \pm 0.5$~K, respectively.
Using the relation between \Tc~and the nominal K concentration $x$ reported in ref.~\onlinecite{Avci2012} 
and wavelength-dispersive x-ray spectroscopy,\cite{Tanatar_2014_PRB}
we obtain $x = 0.22$, 0.24 and 0.28, respectively.
These $x$ values are also consistent with the measured
antiferromagnetic ordering temperature \TN~(which 
coincides with the structural transition from tetragonal to orthorhombic),~\citep{Avci2012}
equal to 
$91 \pm 5$~K and $79 \pm 2$~K, respectively for the two lower dopings.
The sample with $x = 0.28$ shows no magnetic or structural transition.
The resistivity at room temperature of all samples lies between 250 and 350 $\mu \Omega$~cm, 
in agreement with previous studies. \cite{Liu_2014_PRB}
As before,\cite{Hassinger_2012_PRB}
we have normalized the resistivity at $T = 300$~\,K to 300 $\mu \Omega$cm.
Hydrostatic pressures up to 
2.75~GPa were applied with a hybrid piston-cylinder cell,\citep{walker_nonmagnetic_1999}
using a 50:50 mixture of n-pentane:isopentane.\citep{duncan_high_2010}
The pressure was measured via the superconducting transition of a lead wire inside the pressure cell. 
The electrical resistivity $\rho$ was measured for a current in the basal plane of the orthorhombic crystal structure,
with a standard four-point technique using a Lakeshore ac-resistance bridge. 
The transition temperatures are defined as follows: 
\Tc~is where $\rho=0$; 
\TN~and $T_0$ are detected as extrema in the derivative $d\rho/dT$.
%


{\it Resistivity.--}
Fig.~1 shows the in-plane resistivity (top panels) and its temperature derivative (bottom panels) of each sample, for a selection of pressures.
\TN~is detected as a peak in the derivative for the first sample at ambient pressure, and then as a dip for higher pressures or doping. 
The transition at \To~shows up as a sharp peak, below \TN.
In Fig.~2, the full set of derivative curves is displayed for $x = 0.22$ and $x = 0.24$, allowing to track
the anomalies at \TN~and \To~as a function of pressure.


\begin{figure}[]
\includegraphics[width=8.1cm]{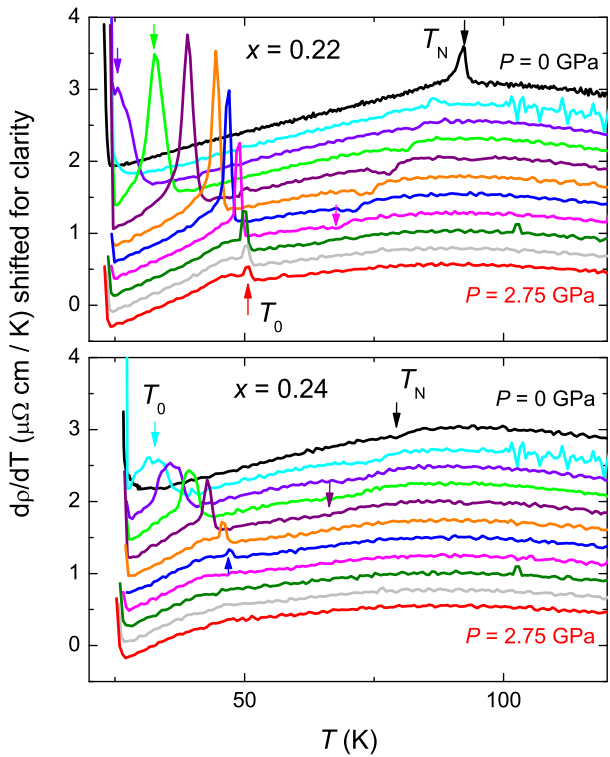}
\caption{
{\it Top}:
Temperature derivative of the resistivity of \KBa~with $x = 0.22$, for 11 different pressures, from
ambient pressure ($P = 0$) at the top (black) to $P~=~2.75$~GPa at the bottom (red), 
with the following intermediate values:
$P = 0.28$, 0.48, 0.78, 0.94, 1.37, 1.68, 2.0, 2.31, and 2.4 GPa.
The curves are shifted for clarity.
The black down-pointing arrow marks \TN~at $P = 0$.
The next down-pointing arrow marks \TN~at the highest pressure where it can still be detected.
\To~shows up as a peak at low temperature.
The up-pointing arrows mark \To~at the highest pressure where the peak can still be detected.
{\it Bottom}: 
The same for $x = 0.24$.
}
\label{Tc}
\end{figure}


As previously reported for samples with lower doping, \citep{Hassinger_2012_PRB}
\TN~decreases linearly with pressure. For $x = 0.22$, the peak in the derivative at \TN~evolves into a dip at 0.48\,GPa.
We are able to follow this dip up to $P = 2$\,GPa, above which it disappears. 
The evolution of the peak at \To~is different. At 0.48\,GPa, the peak at \To~appears. \To~goes up with pressure until it stays almost constant above 2.3\,GPa. 
The height of the sharp peak at \To~increases slightly 
at first, and then
 decreases above $P \simeq 1.5$~GPa.
The behavior for $x = 0.24$ is similar, but
shifted to lower pressures. 
\TN~can be followed only up to 0.94\,GPa. 
The transition at \To~appears as a peak as soon as we apply pressure. 
In fact, a slight upturn of the derivative with decreasing $T$, indicative of an onset of the transition at \To, can be seen even at ambient pressure. 
The peak at \To~stays 
sharp but its height decreases above $P \simeq 1$~GPa, and
the last pressure where it is observed is 1.68\,GPa. 
The curve at this pressure looks very much like the one at the highest pressure in the $x = 0.22$ sample. 
%


{\it Temperature-pressure phase diagram.--}
Fig.~3 presents the temperature-pressure phase diagram for the three samples. 
\TN~decreases linearly with $P$, with a slightly steeper slope at $x = 0.24$. 
By contrast, \To~rises rapidly, at least initially.
At $x = 0.22$, \To~saturates above $P = 2.3$~GPa.
At $x = 0.24$, we can no longer detect \To~above $P = 1.68$~GPa (Fig.~2),
the pressure at which it merges with the \To~line at $x = 0.22$ (Fig.~3).

At $x = 0.24$, the phase diagram is such that if 
the \To~line (blue) saturates at high pressure as it does in the case of $x = 0.22$ (red \To~line),
then a linear extension of the \TN~line (blue) will hit that \To~line,
implying that the t-AF phase would persist to pressures beyond the end of the o-AF phase.

As for superconductivity, note that 
\Tc~decreases as soon as the tetragonal phase appears (Fig.~3),
as found in 
prior studies of Ba$_{1-x}$K$_x$Fe$_2$As$_2$ (refs.~\onlinecite{Hassinger_2012_PRB, Budko_2013_PRB})
and Ba$_{1-x}$Na$_x$Fe$_2$As$_2$, \cite{Budko_2014_PRB, Budko_2013_PRB}
in agreement with the negative $dT_{\rm c}/dP$ expected from the Ehrenfest relation applied to the thermodynamic data.\cite{Boehmer_2015_NatCom}

%
{\it Temperature-concentration phase diagram.--}
Combining our present results with those of our previous study,\citep{Hassinger_2012_PRB}
we plot the temperature-concentration phase diagram of \KBa~in Fig.~4.
For comparison, we also reproduce the phase diagram at zero pressure reported in ref.~\onlinecite{Boehmer_2015_NatCom};
the agreement with our own ambient-pressure data is excellent.
We see that the \TN~line moves down with pressure, in parallel fashion.
This suggests that the critical concentration $x_{\rm N}$ where \TN~goes to zero shifts down with pressure.

On the backdrop of this shrinking o-AF phase,
the tetragonal magnetic phase undergoes a major expansion with pressure
(Fig.~4). 
While the t-AF phase occupies a small area below \TN~at ambient pressure,
its area grows by 
an order of magnitude at $P = 2.4$~GPa.
In other words, at high pressure the tetragonal phase becomes the dominant magnetic phase
in the temperature-concentration phase diagram of \KBa.
In the context of recent calculations, it may be that pressure favours the t-AF phase because it changes the ellipticity of 
the electron pockets in the Fermi surface of \KBa. \cite{Wang_PRB_2015}


\begin{figure}[b]
\centering
\includegraphics[width=8.3cm]{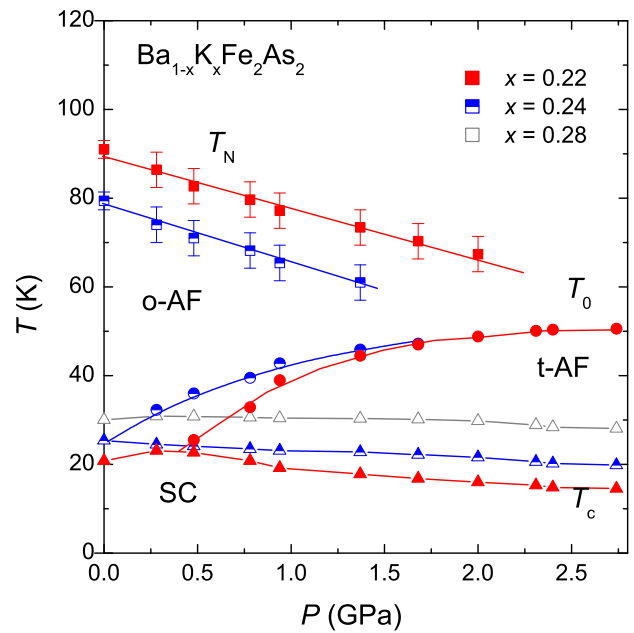}
\caption{
Temperature-pressure phase diagram of Ba$_{1-x}$K$_{x}$Fe$_2$As$_2$, for $x = 0.22, 0.24$ and 0.28
(full, half-full and empty symbols, respectively),
showing the orthorhombic antiferromagnetic (o-AF) transition temperature \TN~(squares), 
the superconducting (SC) transition temperature \Tc~(triangles),
and the tetragonal antiferromagnetic (t-AF) transition temperature \To~(circles).
}
\label{phasediag}
\end{figure}





{\it Summary.--}
In summary, we have shown that the new phase discovered in \KBa~from sharp signatures in the resistivity under pressure\citep{Hassinger_2012_PRB} 
is the tetragonal antiferromagnetic phase observed and identified subsequently by various probes in both  
Ba$_{1-x}$Na$_x$Fe$_2$As$_2$ (refs.~\onlinecite{Avci_2014_NatCom,Wasser_2015_PRB}) and \KBa.\cite{Boehmer_2015_NatCom,Mallet_2015_PRL,Allred_2015_PRB}
Under pressure, this t-AF phase expands enormously, by an order of magnitude for 2.4~GPa in terms of the area it occupies
in the temperature-concentration phase diagram, relative to the orthorhombic stripe-like AF phase that dominates at ambient pressure.
As a result, at high pressure superconductivity exists on the border of a dominant tetragonal magnetic phase.
It is then likely that fluctuations of that double-$Q$ phase play a role in the pairing.
Recent calculations suggest that such fluctuations could actually enhance \Tc. \cite{Fernandes_arXiv_2015}

%
The work at Sherbrooke was supported by a 
Canada Research Chair,
the Canadian Institute for Advanced Research, 
the National Science and Engineering Research Council of Canada, 
the Fonds de recherche du Qu\'ebec - Nature et Technologies, 
and the Canada Foundation for Innovation.
The work at the Ames Laboratory was supported by the DOE-Basic Energy Sciences under Contract No. DE-AC02-07CH11358.
The work in China was supported by NSFC and the MOST of China (\#2011CBA00100).


\begin{figure*}[t]
\centering
\includegraphics[width=17.9cm]{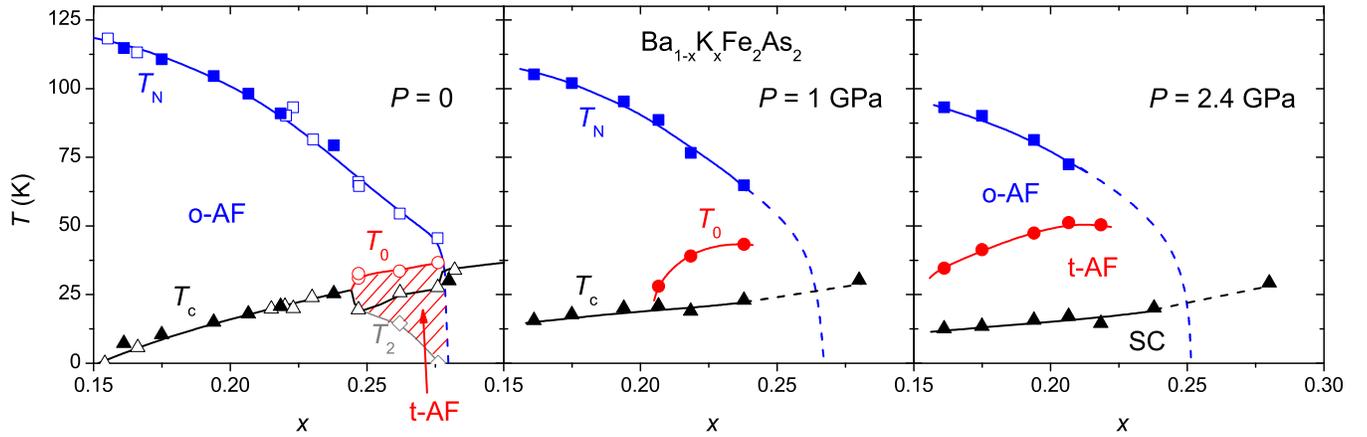}
\caption{
Temperature-concentration phase diagram of \KBa, showing \TN~(blue squares), \To~(red circles) and \Tc~(black triangles),
for three different values of the applied pressure: $P = 0$ (left panel), 1.0~GPa (middle panel) and 2.4~GPa (right panel).
This includes data from our previous study.\cite{Hassinger_2012_PRB}
Ambient-pressure data from ref.~\onlinecite{Boehmer_2015_NatCom} are also shown in the left panel (open symbols),
including a transition back to the o-AF phase, below $T_2$ (diamonds).
All lines are a guide to the eye.
The evolution from left to right, with increasing pressure, reveals a major expansion of the tetragonal magnetic phase (t-AF), on the backdrop of a shrinking stripe phase (o-AF). Extrapolating to higher pressure, we expect the former to become the dominant magnetic phase coexisting with superconductivity in \KBa.
}
\label{dopingphasediag}
\end{figure*}



\end{document}